# The first applications of novel gaseous detectors for UV visualization

J-M. Bidault[1], I. Crotty[2], A. Di Mauro[2], P. Martinengo[2], P. Fonte[3], F. Galy[3], V. Peskov[2],

I. Rodionov[4], O. Zanette[1]

[1]Pole University Leonard de Vinci, Paris, France
[2] CERN, Geneva, Switzerland
[3] LIP, Coimbra, Portugal,
[4] Reagent Research Center, Moscow, Russia

**Abstract**

We have demonstrated experimentally that recently developed gaseous detectors with resistive electrodes combined with solid or gaseous photo-cathodes have exceptionally low noise and high quantum efficiency for UV photons while being solar blind. For this reason they can be used for the detection of weak UV sources in daylight conditions. These detectors are extremely robust, can operate in poor gas conditions and are cheap. We present the first results of their applications to hyper-spectroscopy and flame detection in daylight conditions.

## I. Introduction

In the last few years, a great interest arose for solar blind UV imaging detectors as requested by new applications such as hyperspectroscopy [1-3] and visualization of UV sources in daylight conditions [4].

Currently, advanced microchannel plate detectors (MCPs), sensitive in the spectral interval 300-850 nm [1], are used for these applications. With this approach the very weak UV light has to be detected against the very strong visible light background. This puts very strong constrains on the signal to noise ratio [1, 4] and the development of low noise MCPs is rapidly progressing [5]. However, the recently developed photosensitive micropattern gaseous detectors (see [6, 7] and references therein) may successfully compete with MCPs in some applications. Indeed, these detectors have several interesting features: they are solar blind, they have good position resolution and high efficiency for UV detection. Their planar geometry is very well suited to cover very large area because of the weak constrains on the window size. Last but not least, photosensitive gaseous detectors have extremely low thermo-electronic noise. There are two fundamental reasons for this:
i) in contrast to MCPs, in gaseous detectors there are no high kinetic energy positive ions which cause the so called 'kinetic electron extraction" from the cathode [8], ii) in some gas mixtures, adsorbed layers can be formed on the cathode (see for example [9]) preventing low energy electrons (for example, thermo- electrons) to reach the detector sensitive volume (see below for details).

Recently, we improved the performance of micropattern gaseous detectors by coating the electrodes with a resistive layer which greatly increases their robustness against discharges. Two devices were developed and tested: a microgap resistive plate chamber (MGRPC) [10] combined with a CsI photo-cathode and a thick GEM with resistive electrodes (RETGEM) [11] coupled to a CsTe/CsI photo-cathode [3].

In this work we present the results from the first application of such detectors to the detection of UV photons in daylight conditions.

## II. Application of photosensitive MGRPC to UV hyper-spectroscopy

Hyper-spectroscopy is a new method of surface imaging which provides simultaneously both high position and spectral resolution allowing for remotely studying the chemical composition of the surfaces [1, 3]. This is achieved by means of a special spectrograph (called a hyper-spectrograph, see Fig.1) combined with an optical system which selects a narrow strip (with length A and width $\Delta B$) on the surface under study (hereafter called "strip of interest", see Fig. 2). The hyper- spectrograph forms in its focal plane a reduced image of the strip of size (A/M) x($\Delta B$/M) (where M is a coefficient determined by the optical system) for each wavelength interval. As an example, let's assume that a part of the surface of the strip of interest with coordinates $y_1<y<y_6$ emits a narrow line centered at the wavelength $\lambda_1$, the other part of its surface with coordinates $y_3<y<y_4$ emits a line with wavelength $\lambda_2$ and the other part of the surface $y_2<y<y_5$ emits at wavelength $\lambda_3$. In this case the image of the strip in the focal plane of the spectrograph will look as in Fig 3. For simplicity let's also assume that the two -dimensional (2D) position resolution of the system is $\sim\Delta B$ and the input slit of the hyper- spectrograph is adjusted to the value $\Delta B$/M. In this case in the focal plane of the hyper- spectrograph a spectrally resolved image of the strip with a spectral resolution $\Delta S= S \Delta B/M$ will be formed, where S is the spectral resolving power of the spectrograph $S=\Delta\lambda/\Delta L$, L being the length of the region in the focal plane of the spectrometer on which the spectrum is projected. To record the image a 2D position sensitive photo-detector is needed (see Fig.1) with a typical requirement for position resolution being 50-100 μm (usually M>>1 so a very high position resolution detector is needed for the surface analysis). Often the hyper- spectrograph is installed on an aircraft to perform a scan of the surface of the earth; the strip of interest will move with the aircraft. In such a way a two dimensional image of the surface with both high position and spectral resolutions is obtained. This results in a very high recognition power since such a method provides much larger amount of information than, for instance, a color pictures or the human eye; it allows for quantitative statements about the chemical composition of the surface. Another important application is the detection of sub-pixel objects, i.e. objects with size smaller than the pixel resolution ($\Delta B$/M in the case of our particular example) and the estimation of the relative abundance of the different compounds present inside the pixel. Until recently, hyper spectroscopy measurements were performed in the spectral region 300-860 nm [1] and in most cases using sunlight as the radiation source.

Our work aims at extending the hyper-spectroscopic method into the UV region, namely in the wavelength range 185-230 nm, where the additional information is present. Sunlight in the interval 185-280 nm is fully absorbed in the upper layers of the atmosphere by the ozone, however the atmosphere is transparent in this range at the ground level. More precisely, it is transparent for distances smaller than 100 m and transparent in the range 250-280 nm for larger distances (of the order of a km)

{12,13,14]. Thus to perform the hyper-spectroscopy measurements in the interval of 185-280 nm, an artificial UV source is required which, in most practical cases, has a rather low intensity calling for low noise, high efficiency detectors. In this work we investigate the potential of MGRPC, developed earlier for RICH applications [10], and RETGEM [3, 11] to hyper-spectroscopy.

Fig.4 shows a schematic view of our apparatus. It consists of a hyper-spectrometer combined with a gaseous detector, MGRPC with CsI photocathode or RETGEM with the CsTe/CsI photocathode. The optical system projects the image of the sample under study into the entrance slit of the hyper-spectrometer. Since the detectors used in this work had only 1D position capability (see [10]) the measurements were performed in two steps: first the spectrum at different positions along the strip of interest (width 0.3 mm) was measured, then the detector was turned by 90° (along the line perpendicular to the focal plane) and the image of the strip was recorded for each spectral interval. As an example Fig. 5a shows a 1D digital hyper-spectroscopy image of two superimposed yellow papers (see [3]). During this measurement the gas gain was $10^5$ and the noise rate was ~1 Hz (see Figures 5b-d). The detector was operated without degradation of the quality of the image up to gains of $5x10^5$, while the noise rate in this condition was less than 5 Hz. We believe that it will be impossible to achieve such an image quality with presently available MCPs, as they have a typical noise (after optimization of the gain and the threshold of the electronics) close to 1 kHz. Indeed, when we used a low noise photomultiplier[1] (PMT) for independent spectroscopic measurements, with a noise level close to ~500 Hz and at a gain of $5x10^5$ (an exit slit was installed in the focal plane of the hyperspectrometer in this case) it was possible to detect only light with λ> 250 nm (lines at λ= 253, 296 and 302 nm); the spectrum with λ smaller than 250 nm was completely buried in the noise (PMT's internal noise and the scattered long wavelength light). When the PMT was cooled (by $LN_2$ vapours) to -40°C, the dark noise remained on the level ~45 Hz (the PM gain was $5x10^5$, the mean signal amplitude after the multiplication was ~300 mV, the threshold was of 100 mV). The noise PM pulses at this temperature were grouped in bunches with time between pulses in bunches varying from a few to 10 μs indicating their ion feedback nature. However, even the cooled PM was not able to detect the radiation in the spectral range of 185-250 nm due to the strong scattered long wavelength light. This demonstrates the advantage of our approach for measurements in the spectral interval of 185-250 nm. Moreover MGRPCs are considerably cheaper than PMTs or MCPs. It is not very difficult to equip ours detectors with an electronic readout allowing for 2D position measurements. MGRPCs detectors are insensitive to vibrations and therefore they can be mounted, together with an UV source, on helicopters and used to perform environmental monitoring of the earth.

**III. The detection of flames**

A second promising application is the detection of UV emitters, like corona discharges, sparks and, for practical reasons, flames under day light conditions. As mentioned above, the sunlight in the wavelength interval 185-280 nm is fully absorbed in the upper layers of the atmosphere, offering a unique possibility to detect artificial UV sources without background from the Sun. Several commercial devices already exploit this feature (see for example [15] and references therein). However, more sophisticated detectors can provide better and more accurate information leading to important technical and commercial applications [4].

---
[1] For example, Slumberger 541F-09-17

Currently, image intensifiers combined with narrow band filters are used in some special applications for UV visualization [4, 16]. The MCPs used in these detectors are sensitive in a wide range of wavelengths: from UV to visible light (they have bi-alkaline photo-cathodes) and to suppress the light with $\lambda>280$ nm, narrow band filters are used with a transmission less than 10% in the wavelength interval 240-280 nm. This implies low efficiency for the detection of the UV light. These devices are difficult to produce, and have high cost; therefore they did not yet find wide application.
We have recently demonstrated that photo-sensitive gaseous detectors with resistive electrodes can compete with image intensifiers in the detection of flames on relatively short distances, less than 100 m[3,15].

The MGRPC with CsI photocathode described above was able to detect the flame from a match or a cigarette lighter at a distance of 30 m inside a fully illuminated building: the counting rate produced by the UV light from these flames was 400-450 Hz, whereas the counting rate in the same conditions but in absence of the flame was of the order of 1 Hz mainly caused by cosmic rays. Even higher sensitivity was achieved with RETGEMS combined with CsTe/CsI photo-cathodes [3] or with RETGEMs combined with SbCsI/CsI [17]. These detectors were able to detect small flames at a distance of 60 m. In these detectors the CsI protective layer did not allow for small energy photo-electrons to pass through [18] and as a result it cuts the sensitivity of the photo-cathode to photons with $\lambda$ larger than 280 nm [19, 20], enhancing the sensitivity to short wavelengths ($\lambda<280$nm) [19, 20]. Because the transmission of this film strongly decreases with the electron's energy, the CsI coating allows also considerably suppress the thermo-electronic noise

Unfortunately, detectors with CsTe/CsI or SbCsI/CsI photo-cathodes are very demanding in term of gas purity in order to achieve long term stable operation.

In this work we developed an improved version of solar blind gaseous detectors using photosensitive vapors EF [21] or TMAE [21] instead of solid photocathodes. The main improvement is the capability to operate under poor gas conditions, leading to a very competitive production cost.

The scheme of this detector is presented in Fig.6a. It consists of a gas chamber (filled with a mixture of Ar with EF or TMAE vapors at a total pressure of 1atm) combined with a lens (focal length F= 50 cm). Two RETGEM, operated in cascade mode, and a readout plate are installed inside the gas chamber.

The RETGEM has an active area of 40x40 mm$^2$ or 100x100 mm$^2$ and a thickness of 0.4 mm. The diameter of the holes is 0.3 mm, the pitch 0.7 mm. Instead of graphite coating in this work we used either CrO or CuO resistive coating [17], which, compared to graphite coating, allowed for more stable operation of the RETGEM because it does not absorb photosensitive vapors. The schematic drawing of the readout plate is shown in Fig. 6b. It was made out of a G-10 plate with a fan-out of 10 Cu strips converging in one point corresponding to the centre of the lens placed 50 cm away from the middle of the RETGEM plane (see Fig. 6b). Each strip was individually readout.

In this geometry (F>>d, where d is the size of the RETGEM), the UV photons from the flame are projected along one or two strips and cause the photo-ionization of the photo-sensitive the vapors. The photo-electrons created in the drift region trigger Townsend avalanches in the holes of the RETGEM which in turn induce pulses on the readout strips. The typical overall gain at which we operated the double step RETGEMs was ~$10^5$. Since photosensitive vapors have a zero efficiency for λ larger than 220 nm [21] such a detector was able to operate with a very small background counting rate even in presence of direct sunlight (see Fig. 7a). Also note that at higher gain the pulse height spectra began to deviate from the exponential form, see Fig. 7b. By pulse counting it is possible to obtain a 1D digital image of the flame. As an example Fig.8 a, b shows the image of an alcohol flame, approximately 5x5x5 cm$^3$ placed 70 m away from the detector. The image was obtained in open air on a sunny day. Note that for the human eye such a small flame is practically invisible in the same light conditions. The comparison between Figures 8a and 8b shows that the sensitivity of the detector filled with the TMAE vapors was a few time higher than with the EF vapors. However, EF vapors are not chemically aggressive and remain stable even in air.

During the test with the EF vapors we discovered that RETGEM could operate at gains of $3x10^4$-$10^5$ in gas mixtures with air and even in pure air. For example, the detector could operate perfectly well with the mixture of Ar+EF+ 10% of air. This offers a very simple and cheap design in some applications since no special tightness of the gas vessel is required. In a particular case of the UV photon detection, the efficiency remained high only at small air concentration (<1%). Note that even if the EF vapors leak into the room, they are not toxic or dangerous for the environment. A drop of liquid EF placed inside the detector will make it able to operate even in presence of small leaks because the leaking out vapors will be compensated by the evaporation of the EF drop. Of course the voltage at which detector operates will gradually changed, but it could be easily compensated by a simple automatic HV circuit. Certainly we do not suggest in this work to manufacture leaking detectors, what we want to emphasize is that the design of this detector could be very simple and thus cheap.

We compared the sensitivity of our gaseous detector filled with EF to a commercial UV device. The result was that our detector was about 100 times more sensitive than commercial UV sensors of flames, for example Hamamatsu R2868 [22], see Table 1.

We also tried to compare our detector with the MCP- based UV image intensifier; it had a bi-alkali photocathode combined with an interference UV filter. Due to the high rate of noise pulses (>1 kHz with the electronics used for this detector) we were able to detect flames from matches at a distance of <30 m. Taking into account the difference in the focusing optical system used for the image intensifier and for our detector, we concluded that our detector was 5-7 times more sensitive at short distances (less than 100 m). However, for the flames placed at a distance larger than 150 m, the sensitivity of our detector began degrading due to the abortion of the light in air and the MCP becomes much more e sensitive since it operates in the wavelength interval of 240-280nm. Slightly better results were obtained with TMAE filled detectors which were more sensitive than the image intensifier up to a distance of 200 m.

**IV. Discussion:**

Due to the low noise and the high sensitivity to UV and imaging capabilities, our detectors offer an attractive solution for the detection of weak UV radiation sources in the presence of a strong visible background. For example, they are almost 100 times more sensitive than commercial UV flame detectors.

Low noise, high sensitivity PMTs exist on the market since long time and low noise MCPs are under development. However, these devices are, for the moment, too expensive for massive use in flame detection. Moreover PMTs, due to the resistive chain of the voltage divider, consume a lot of power while an ideal flame detector should be battery fed. Gaseous detectors offer important advantages: they are cheap, have high efficiency to UV light [21] and could be battery fed. Gaseous detectors operating with photosensitive vapors have additional advantages: they have practically zero efficiency for λ larger than 225 nm whereas most of the detectors with solid photocathodes may be noisy in the presence of the sunlight due to the small sensitivity to long wavelength radiations (see [2] for more details). Thus after long operation in direct sunlight conditions the photocathode may degrade due to the aging effect caused by visible light. In the case of the hyper-spectroscopy application, photosensitive gaseous detectors can compete with MCPs only for wavelength shorter than 250 nm and only in measurements performed in the presence of a strong visible background. Low noise MCPs with CsTe or CsI photo-cathodes today under development may offer in the future better performance than detectors.

## IV. Conclusions

Our studies have shown that gaseous detectors with resistive electrodes can compete with commercially available UV sensitive detectors in several applications, namely for hyper-spectroscopy and flame visualization up to a distanced of about 200 m. At the same time compared to commercial UV flame sensors or UV image intensifiers they can offer much larger sensitive area at lower price.

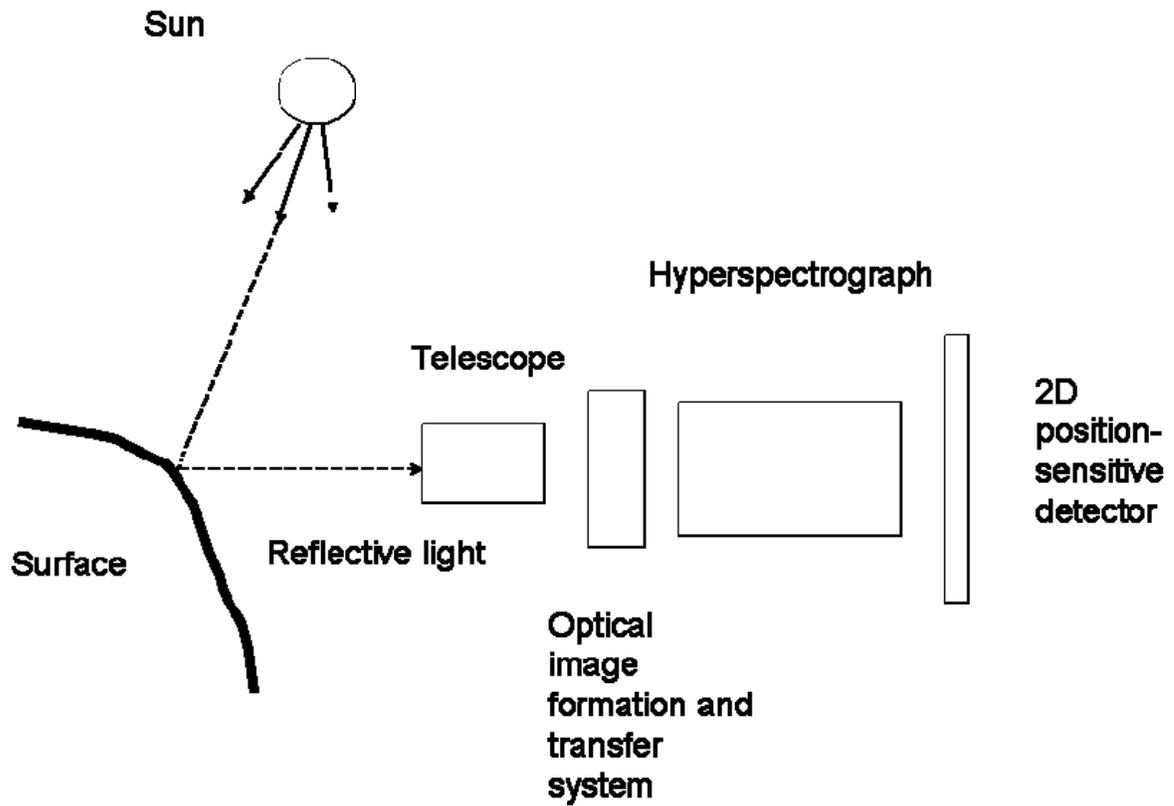

Fig.1. Scheme of principle of a device for hyper-spectroscopy image taking. An optical image formation and transfer system allows one to select a strip if "interest" on the investigated surface. The spectrometer forms in its focal plane a2D image: in one coordinate a 1D image of the strip in the given spectral interval and in the other coordinate its spectra

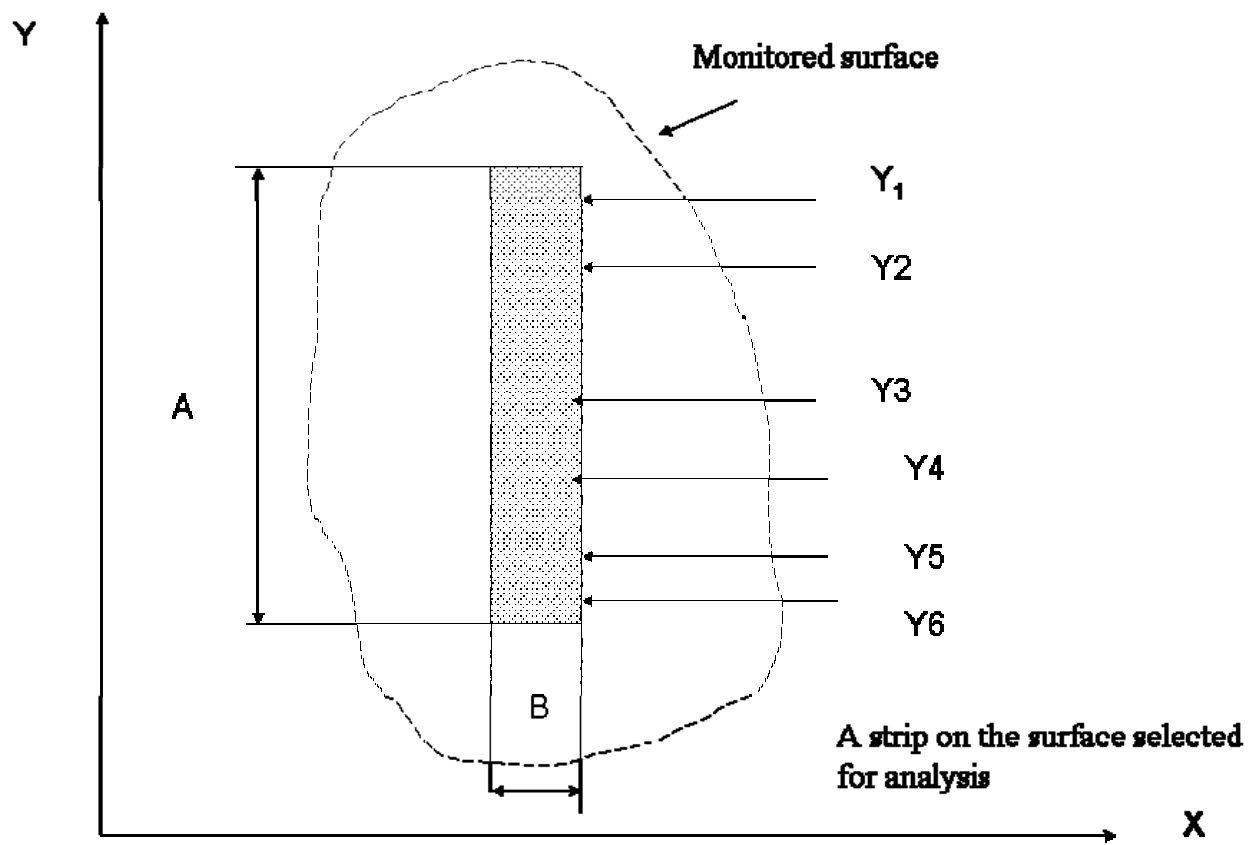

Fig. 2 A simplified picture of the strip selected by the optical system on the investigated surface (a "strip of interest").

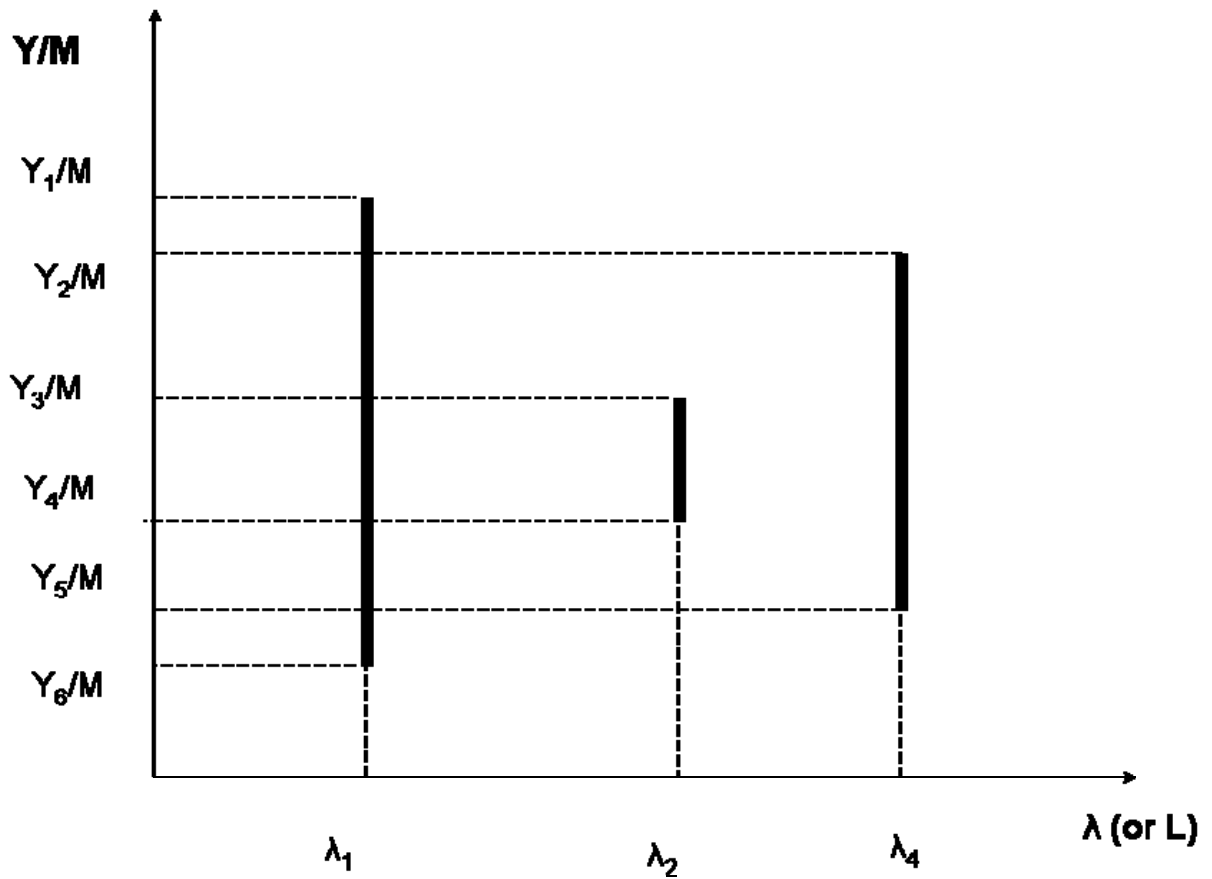

Fig. 3. A simplified example of a two dimensional image of the selected strip on the investigated surface formed in the focal plane of the hyper spectrograph (see [1] for more details). In this particular example some a part of the selected surface (strip) $y_1<y<y_6$ emits lines at wavelength of $\lambda_1$, $\lambda_2$, $\lambda_3$ correspondingly. This example demonstrates that the y coordinate shows the image of the strip in the given spectral interval and the L-coordinate shows the spectra. By the scanning of the investigated surface (moving the strip of the interest in the direction perpendicular to its length) one can measure the spectra of the surface in each selected point /pixel on the surface.

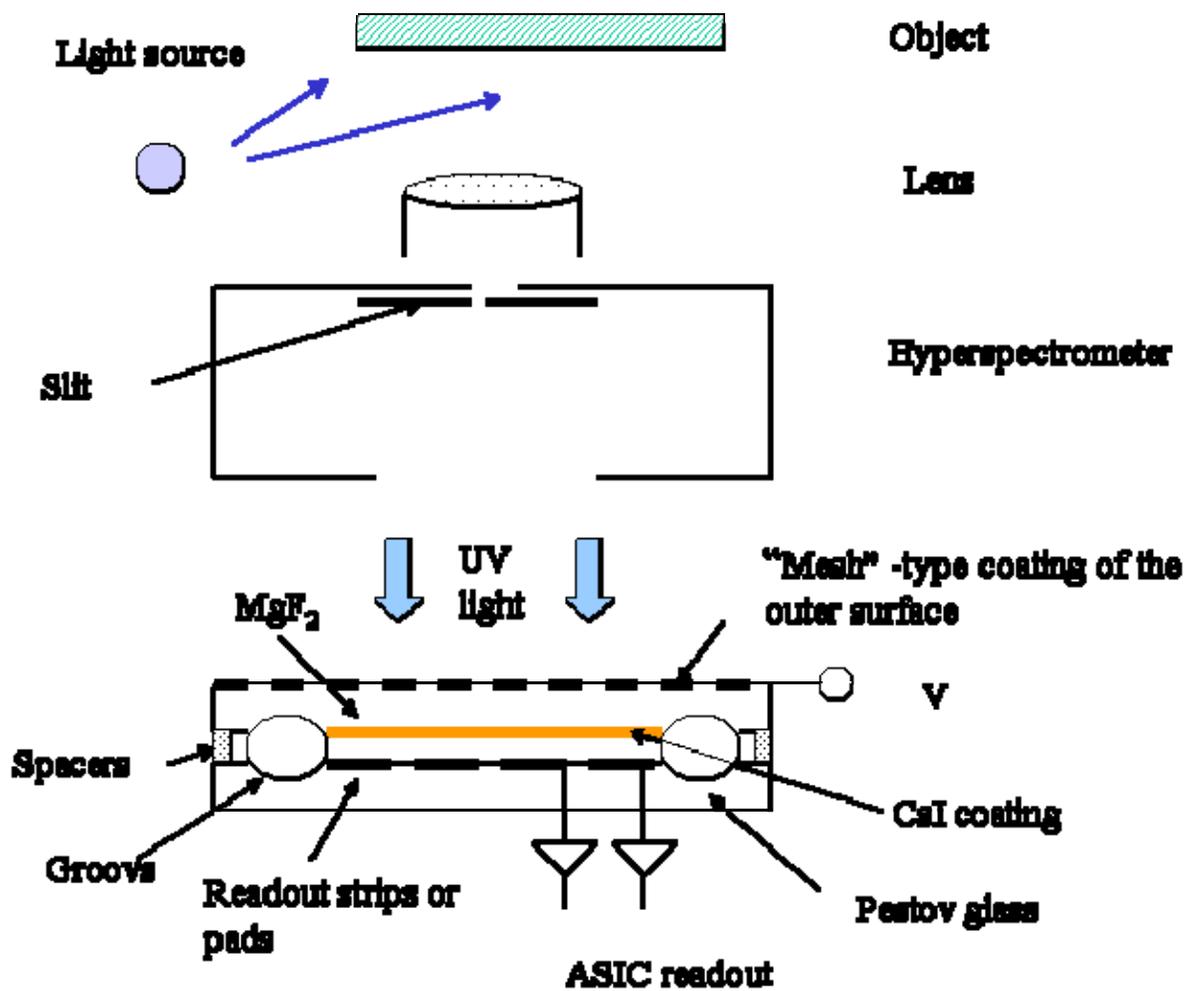

Fig. 4. A schematic drawing of the hyper-spectrometer combined with photosensitive MGRPC

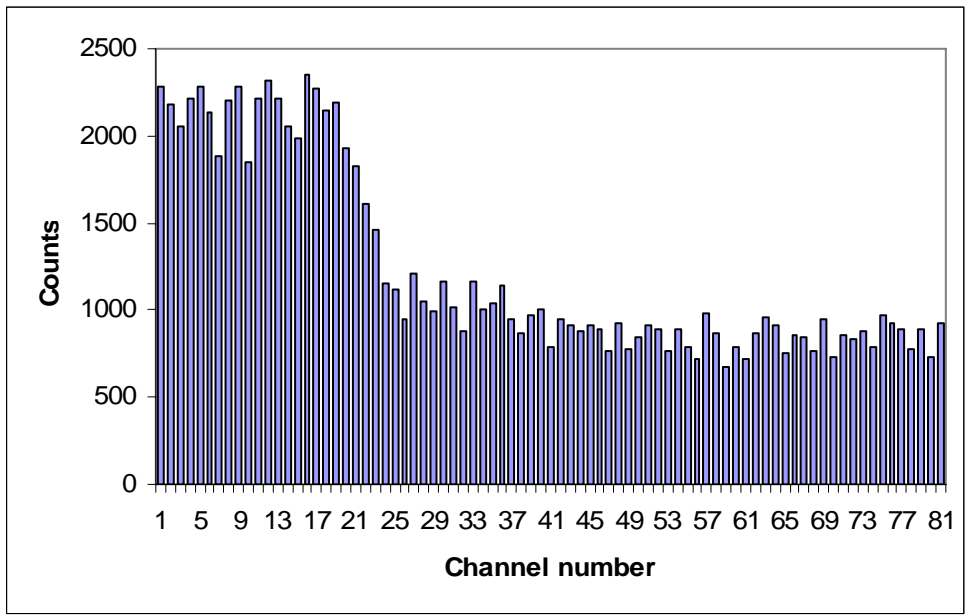

Fig. 5a. 1D digital visualization of the boarder between two yellow papers at λ= 194 nm. In these measurements the mean amplitude of the pulses (with a typical for a single electron detection exponential shape) was 175 mV, the ASIC threshold was of 50 mV. The gas gain ~$10^5$. The number of counts was accumulated during 1s

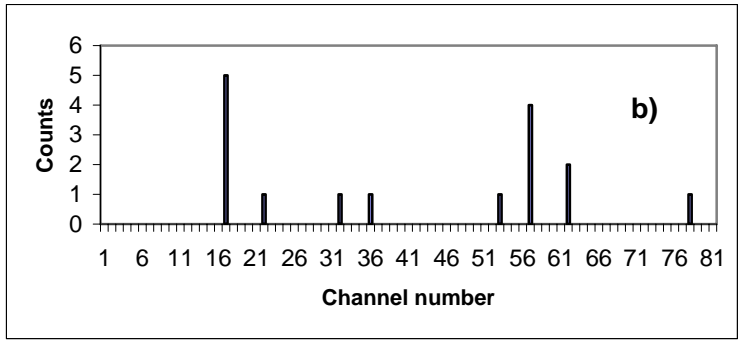

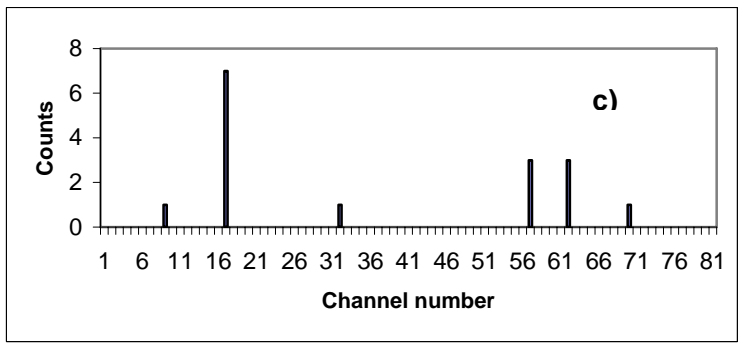

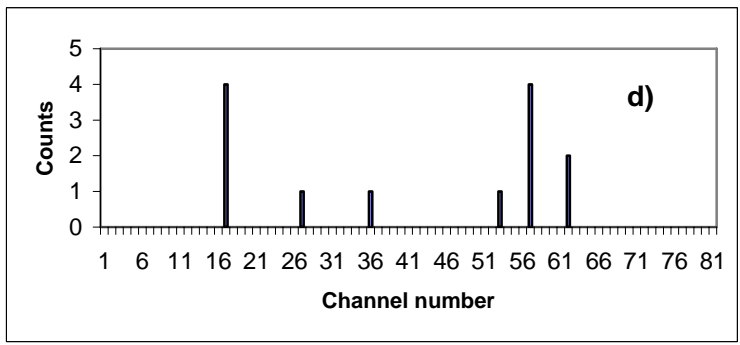

Fig 5b-d. Examples of consequent measurements of the detector noise when the Hg lamp was switched off. In each measurement the noise pulses were accumulated during 10sec. One can see that there are a few particular channels which give the most contribution to the noise. These particular channels were noisy only at gas gains of $10^5$ and higher so we attribute them to the electron emission from some fixed sports on the CsI photocathode

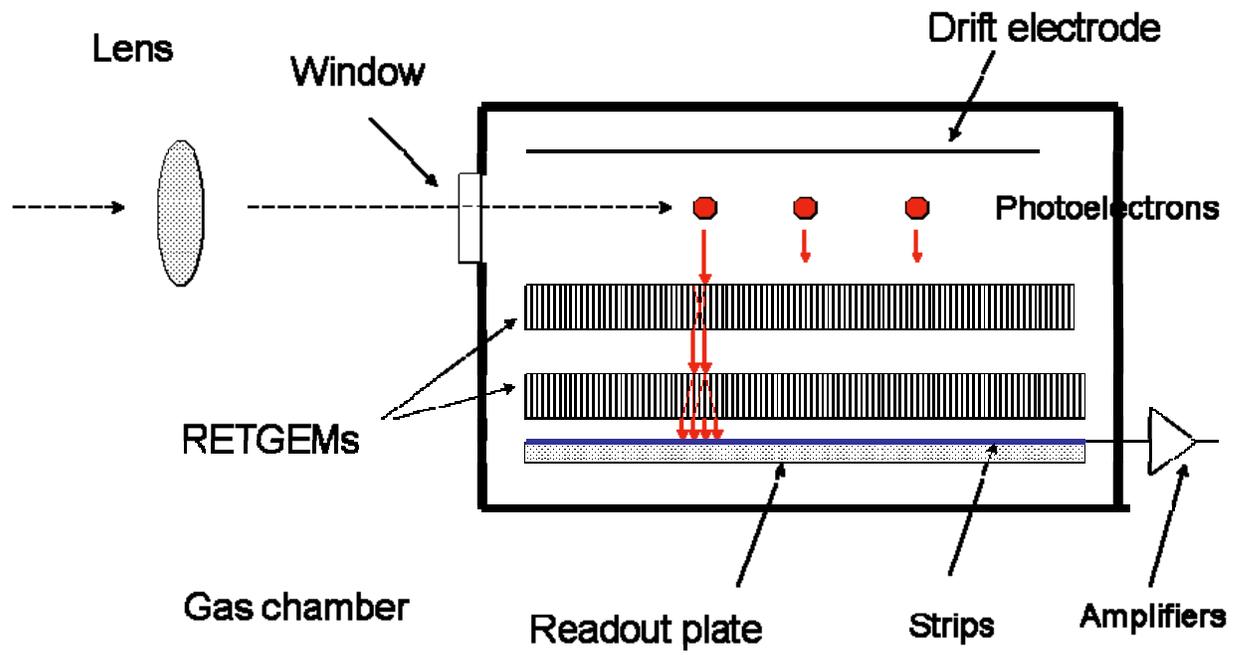

Fig. 6a. A schematic drawing of the RETGEM for visualization of flames.

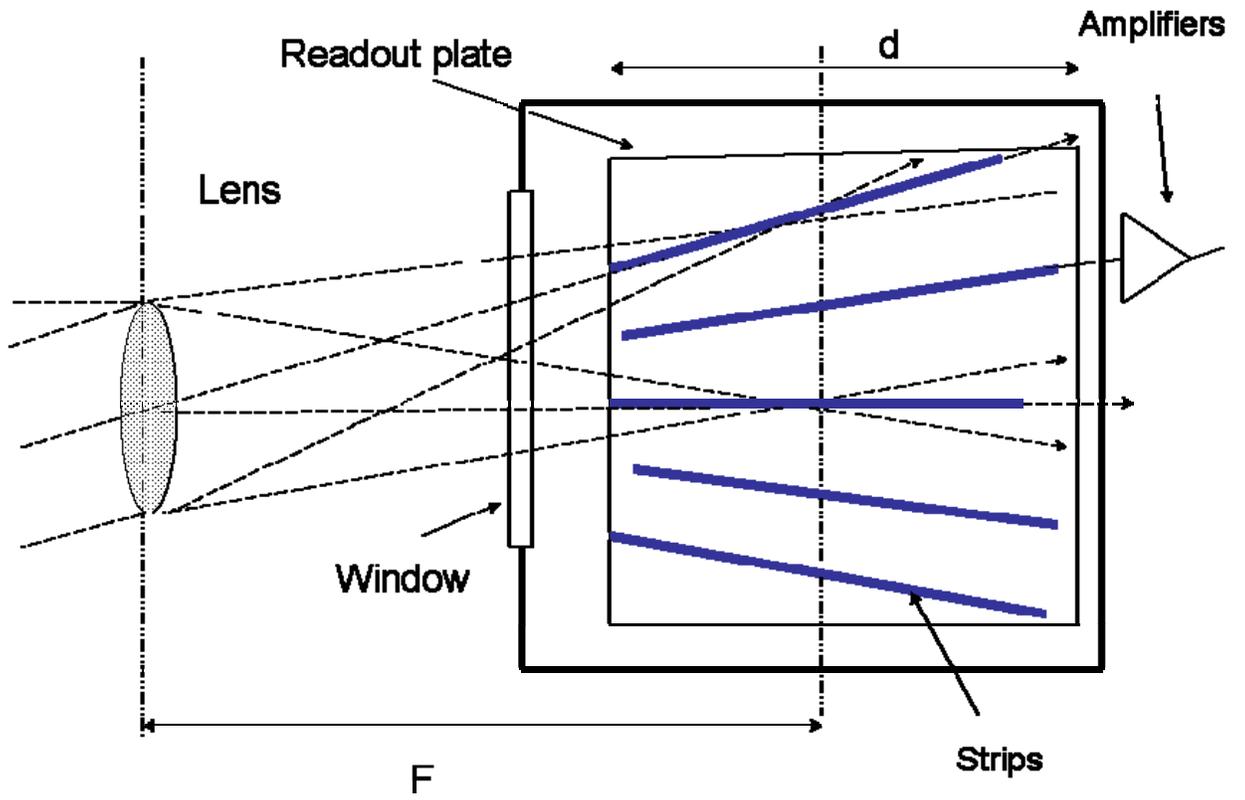

Fig.6b. A schematic drawing of the readout plate. For simplicity only six readout strips and one amplifier are shown.

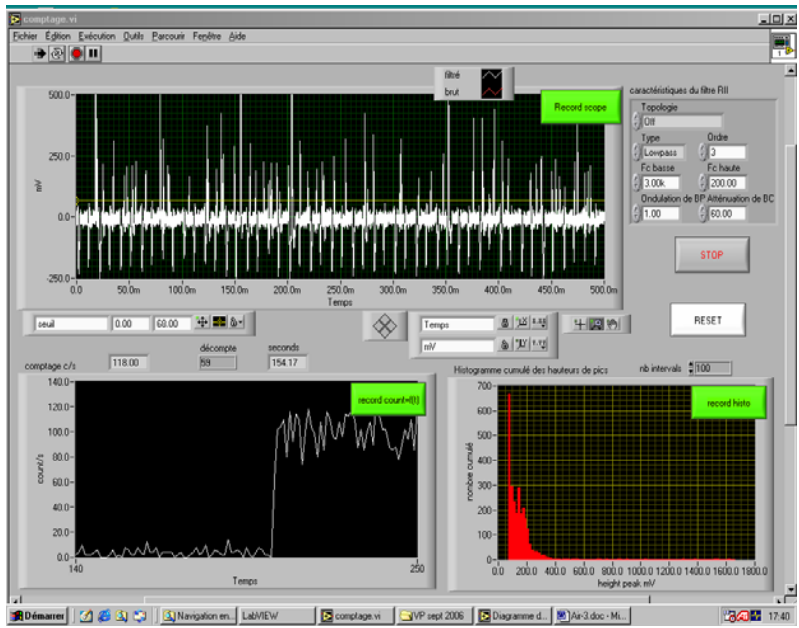

Fig 7a. An example of analog signals from the double RETGEM (all strips are connected to a single amplifier) operating at a gain of $3\times10^5$ and detecting the signal of a few Hz due to the direct sunlight hitting its window (a background caused by the surface photo-effect from the chamber's electrodes: the drift electrode and the cathode of the first RETGEM).
In the middle of measurements a flame at a distance of 7 m was ignited and caused the counting rate of ~110Hz.

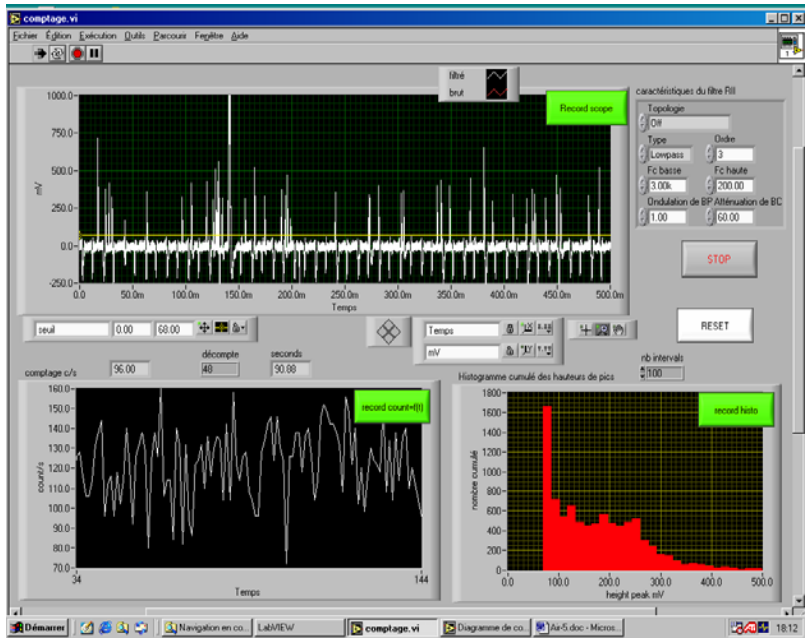

Fig 7b. A copy of the Labview screen showing counting rate and pulse height spectra at gain of close to $10^6$.

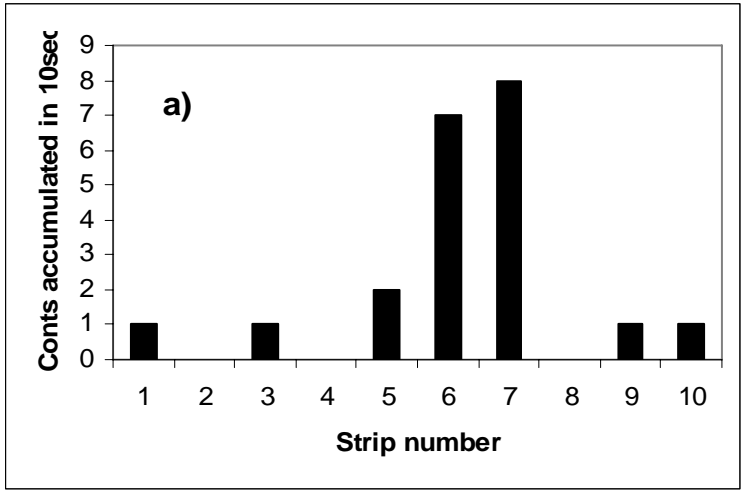

Fig.8a. A digital image of an alcohol flame ~5x5x5 cm$^3$ obtained at a distance of 70 m outside the building and in the presence of full sunlight. The detector is filled by the mixture of Ar+EF vapor at a total pressure of 1atm.

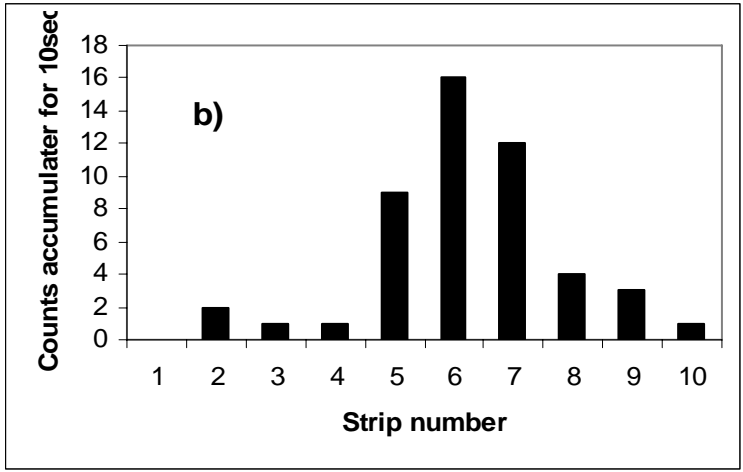

Fig.8b.The digital image of the flame obtained almost at the same conditions with a detector filled with Ar+TMAE vapor at a total pressure of 1 atm.

| Hamamatsu R2868 | | Our detector |
|---|---|---|
| Distance (m) | Mean number of counts per 10sec | Mean number of counts per 10sec |
| 1 | | |
| 1,1 | 583 | |
| 2,5 | 99 | |
| 3 | 76 | 6376 |
| 4,5 | 28 | 3166 |
| 7 | | 1024 |
| 10 | 6 | 518 |
| 20 | | |
| 30 | 0.1 | 63 |

Table 1. Counting rate from the commercial UV flame sensor Hamamatsu R2868 and from our detector both detecting the UV light from a candle placed at various distances from the detectors under indirect sunlight conditions. The mean amplitude of the signal from our detector was 500 mV (exponential shape), the mean amplitude from the R2868 after the attenuation by the probe (1:10) was of 2 V (narrow pick shape), the threshold of the scalier was set to 200 mV. The counting rate at the same conditions but without the candle was ~1 Hz for the R2868 and a few Hz for our detector.